# Phenomenological theory of photo-magneto-electric effects
# Application to the directional photovoltaic effect in LiNbO$_3$

B. Mettout and P. Gisse

*PSC. Université de Picardie, 33 rue saint Leu, 80000 Amiens, France.*

We propose a phenomenological theory working out light illumination effects on the equilibrium values of any macroscopic crystal tensor (polarization, magnetization, susceptibilities, strain tensor, elastic coefficients…). It also encompasses non-equilibrium light-induced quantities such as an electric current and heat flow. We use a single phenomenological approach based on Wigner spherical functions for predicting symmetry-related photo-induced phenomena, including photovoltaic, photoelectric, photo-magnetic, photorefractive, photo-galvanic, photo-elastic and optic rectification phenomena. For each crystal magnetic point group and each tensor type, response functions are calculated vs. the propagation and polarization directions of the incident electromagnetic beam. Their forms are determined by crystal and light symmetry group interconnection. We pay special attention to time and space reversal symmetries, which play a dominant role in the intricate symmetry breakdowns of multiferroic materials. Photovoltaic and photo-magneto-electric effects in LiNbO$_3$ illustrate our theory.

mettoutb@yahoo.fr
patrick.gisse@u-picardie.fr

## I. INTRODUCTION

Internal electric fields in semiconductor materials illuminated with photons having energies above the band-gap width separate induced electron-hole pairs and generate electric currents along the field direction. The corresponding induced current defines the response function of the *photovoltaic effect* [1] observed in BaTiO$_3$ more than six decades ago [2]. The main response is independent of the beam and polarization directions because it depends only on the density and mobility of light-generated free carriers. Furthermore, to a first approximation it depends on the material symmetry and structure only in terms of the existence of a spontaneous polarization axis and the resistivity tensor anisotropy.

Internal fields can be produced at p-n or Schottky junctions [3] and domain walls [4] or can be induced by the spontaneous polarization in ferroelectric materials [5]. The latter case, first discovered in LiNbO$_3$ in 1969 [6], is denoted as the *bulk photovoltaic effect*. It yields important technological applications in the domain of solar-cell energy converters [7]. The cell efficiency is mainly limited by the solar spectrum vs. band gap compatibility [8]. The recently discovered [9,10] photovoltaic effect in BiFeO$_3$ and KBiFe$_2$O$_5$ has promising features in this respect. Indeed, in the latter compound the energy gap (1.6 eV) is solar compatible and exhibits strong ferroelectric properties up to 780 K. Furthermore, its multiferroic character raises more fundamental questions concerning the connection between ferroelectric and/or magnetic orderings and the photovoltaic response.

Indeed, when one focuses on finer directional wave-induced effects, the complete magnetic point group, together with the wave vector and polarization directions of the incident beam, begins to play non-negligible roles beside the simple isotropic effects depending on beam intensity. The first effect comes from the wave polarization direction, which determines the probability and final states of quantum electronic excitation processes and modulates the induced free charge density [11]. In addition, the wave vector direction has effects on the excited electron gas through the direct electron-photon momentum exchange [12]. This relativistic process, responsible for the classical *photoelectric effect* in metals, yields in principle small contributions to the induced current in insulators. Nevertheless, increasing the beam intensity can (almost) saturate[1] the excited carrier density and, consequently, the polarization-induced part of the current so that the photoelectric part can become observable at sufficiently high intensity.

Phenomenological as well as microscopic theories of the photovoltaic and related effects in insulators have been well established since the works of Belinisher [11,13] and Sturman [1,14] in the 70s. The photovoltaic current is calculated in traditional phenomenological approaches on using power expansions of the induced current $J^j$ vs. static electric applied field $E^i$, wave electric amplitude $e^i$ and wave-vector components $k^i$:

$$J^j = \sigma_{(d)}{}^i{}_j E^j + \sigma^i{}_{jl} E^j E^l + \sigma_{(ph)}{}^i{}_{jlm} E^j e^l e^m + \chi^j{}_{jlm} k^j e^l e^m + \beta^i{}_{jl} e^j e^l + ... , \quad (1)$$

where $\sigma_{(d)}$, $\sigma_{(ph)}$, $\sigma$, $\chi$ and $\beta$ are tensors characteristic of the crystal symmetry[1]. In the absence of an applied field, the current is only determined by the entrainment effect described by the fourth-rank tensor $\chi^j{}_{jlm}$, which comes from momentum transfer between photons and electrons, and by the third-rank photovoltaic tensor $\beta^i{}_{jl}$ arising mainly from the electric force due to the spontaneous polarization of the material. Let us notice that Sturman [1] denotes by *photovoltaic effect* only the current contributions due to the photovoltaic tensor while, in the present article, we will not distinguish a priori photoelectric-type (with momentum transfer) and photovoltaic-type (internal electric forces) contributions. However, a posteriori identification of these two contributions is possible for specific terms in the response functions on inspecting their symmetry properties.

Eq. (1) enables one to describe a large variety of phenomena, taking into account the magnetic point symmetry of the crystal via the form of the tensors $\beta$ and $\chi$. However, it presents some inconveniences that lead us to propose an alternative approach. First, although it is natural to use a series expansion vs. the field component $e^i$, which can be made arbitrarily small by experimental tuning, this is no longer the case for the wave vector components $k^i$, which can in no way be considered as small quantities. In addition, $k^i$ and $e^i$ are not independent variables since the vectors **e** and **k** are perpendicular so that coupled terms in the right-hand side of Eq. (1) become ambiguous. Finally, the series expansion in $k^i$ validity assumes the response function is analytic and non-singular at $k^i=0$ while, in this limit, the field symmetry group undergoes a discontinuous change, incompatible with this non-singular character.

Thus, we prefer to use the beam intensity $I_{ph}$ and the three Euler angles describing the orientation of the electromagnetic frame **e**,**b**,**k** with respect to the laboratory frame as independent wave variables. No power expansion is then possible (except vs. $I_{ph}$) and Eq. (1) must be replaced with an expansion based on Wigner functions [15] $D^L_{mn}(\mu)$ ($\mu$ stands for the set of Euler angles $\mu=\{\alpha,\beta,\gamma\}$), analogous to the well-known expansion in spherical harmonics for vector quantities. The mathematical features of such an expansion are presented in section III. These functions are very convenient for taking into account both crystal and wave symmetries. However, they are less powerful than a series expansion for evaluating the relative magnitudes of the different degrees in the expansion. In Eq. (1), the terms with highest powers give negligible contributions when the field $e^i$ is sufficiently small. Analogously, terms with small $L$ give dominant contributions to Wigner expansions when one considers only the measurable slow angular variations of the current. Unfortunately, one cannot conclude anything about the $I_{ph}$ dependence of the Wigner expansion coefficients, except that they all vanish when $I_{ph} \to 0$.

On the other hand, Wigner expansion ensures precise and easy accounting for crystal and wave symmetries. Indeed, the functions $D^L_{mn}(\mu)$ transform as irreducible representations of both groups. To explain this, the wave symmetry and its symmetry-breaking effects are analyzed in section II. Moreover, the crystal group acts only on the second Wigner index $n$ whereas the wave group acts on the first index $m$. Each generator of the wave group cancels a number of expansion coefficients depending on $\mu$, whereas the crystal generators cancel independent sets of coefficients depending on $n$. Thus, both groups provide an independent restriction of the Wigner expansion form. We denote by *external selection* the corresponding restriction procedure involving the wave symmetry operations and denote by *internal selection* that involving crystal symmetries (details of the corresponding internal and external algebras can be found in Ref.[16], within another physical context). These two procedures, presented in section IV, allow us to write the most general form of the induced current response functions associated with each of the 122 magnetic point groups [16], particularly those of multiferroic materials in which both time and space reversal symmetries are usually lost [18,19]. Another immediate benefit of our approach is the fast, simple and straightforward aspects of the selection calculations.

Excitation of electrons in the conduction band of semiconductor materials can also modify their equilibrium structural or magnetic properties. Our approach applies to response functions associated, in steady states, with any physical quantity described by tensors of any rank $N$ and any symmetry with respect to time, $T$, and space, $I$, reversals. Thus, it permits accounting for a large variety of additional photo-induced phenomena. When $N=1$, we have worked out the response functions associated with (i) a polar vector anti-symmetric with respect to T, which describes the *photovoltaic effect* when the vector is the electric current (section V.A) as well as the *photo-toroidal effect* when the vector is the toroidal moment [22] (section V.B); the response function associated with (ii) a polar vector symmetric with respect to T, which describes the *optical rectification effect* [21] when the vector is the macroscopic polarization, and the response function associated with (iii) an axial vector anti-symmetric with respect to T, which describes the *photo-magnetic effect* [22] when the vector is the macroscopic magnetization (section V.B). When $N=2$, the response functions of the dielectric and magnetic permittivity and of the elastic strain tensor (symmetric with respect to both T and I) describe the *photorefractive effect* [23] involved in non-linear optics and the *inverse photo-elastic effect* [24]. Finally, the linear magnetoelectric tensor (anti-symmetric under *T* and *I*) of the *photo-magneto-electric effect* [25] is worked out in section V.C.

In addition, we present in section V an illustration of our theory with the ferroelectric photovoltaic crystal[6] $LiNbO_3$. Its trigonal group permits an efficient internal procedure yielding simple and pedagogical forms for the response functions. Nevertheless, its non-magnetic character fails to illustrate some of the most complex photo-effects involving intrication of polar and magnetic symmetry breakdowns.

**II. MEAN SYMMETRY OF THE LIGHT BEAM**

Light acts locally as an external generalized field on the crystalline and magnetic structures of illuminated materials. Indeed, the response of the crystal at macroscopic time scale is static since it involves a time average over many periods of the electromagnetic wave. Moreover, when the crystal is crossed by the wave it appears as a continuum since the wavelength is much larger than the lattice spacing. The irradiation effects on the material properties can then be evaluated by considering static and homogeneous tensor quantities in exactly the same way as when the material is submitted to static fields [25]. The onset rules of such tensors follow the same symmetry constraints in both cases and can be approached phenomenologically.

In ordered media the response functions depend on the crystal (or order parameter) point symmetry, together with the symmetry of the mean wave. Let us first analyze the linearly polarized wave symmetry group. *e* and *b* represent the wave amplitudes, *k* its wave vector and $\phi$ its phase (*E* should not be confused with *E* in equation (1)):

*E* = *e* exp i(*k.r*–$\omega$t+$\phi$)
*B* = *b* exp i(*k.r*–$\omega$t+$\phi$)

*e*, *b* and *k* are parallel to an orthonormal frame $\Sigma_\mu = I_1, I_2, I_3$ ($\mu$ represents a set of Euler angles that will be defined precisely hereafter). In contrast to (*e*,*b*,*k*), ($I_1,I_2,I_3$) are polar vectors invariant under time reversal. The mean symmetry group $G_{em}$ of the wave contains at least the group of the triplet *e*,*b*,*k*. The latter is the point group 2'mm' where 2' means the combination of time reversal *T* with a twofold rotation axis 2 around $I_1$, and m' means the combination of *T* with a mirror plane m normal to $I_2$. In addition, after a twofold rotation around *k*, *e* and *b* are transformed into –*e* and –*b*. This can be compensated by a $\pi$ phase shift of $\phi$ in equation (1). Since the mean properties of the wave are independent of the phase, this rotation belongs also to $G_{em}$. One finds finally (*I* stands for space inversion):

$G_{em}$=mmm'={*e*,$C_\pi(I_1)T,C_\pi(I_2)I,C_\pi(I_3)IT$} $\otimes$ {*e*, *IT*},  (3)

which is isomorphic to $D_{2h}$. $C_\pi(I_3)$ stands for the twofold rotation around the axis $I_3$. $G_{em}$ has three generators: *IT*, $C_\pi(I_3)$ and $C_\pi(I_1)T$ or, equivalently, $C_\pi(I_1)T$, $C_\pi(I_2)I$ and $C_\pi(I_3)$.

When the beam travels within a material it breaks its symmetry and, accordingly, it induces the onset of a number of physical quantities. Although $G_{em}$ does not permit the onset of any non-trivial scalar quantity (i.e. anti-symmetric with respect to time and/or space reversals, *I* and *T*), it permits a toroidal vector (anti-symmetric under *I* and *T*) parallel to *k* (e.g. the Poynting vector). The current density has the same symmetry properties; nevertheless no current is present when the beam travels in the vacuum because no electric charge is then available, whereas, in a conductor, an electric current can flow in the direction of the Poynting vector (photoelectric effect). Analogously, $G_{em}$ permits a second-rank tensor symmetric under *I* and *T* (e.g. deformation tensor or dielectric susceptibility) with non-zero diagonal components (in the frame $I_1,I_2,I_3$) and a second-rank tensor anti-symmetric under both *I* and *T* (e.g. magneto-electric susceptibility) with two non-zero components, $T^{13}$ and $T^{31}$. All these effects are present in any material, even in isotropic liquids. For crystalline materials many other effects can take place under light irradiation, depending on the symmetry of the material.

One can also illustrate the wave-induced symmetry-breaking effects by considering the onset of various types of physical vectors when various types of isotropic liquids are illuminated. No calculation is then necessary to predict the directions of these vectors, as summarized in Table I: They are either forbidden by symmetry or directed parallel to *k*. The four types of isotropic liquid are (i) a normal achiral liquid (invariant under *T* and *I*), (ii) a non-racemic mixture of chiral molecules (invariant under T), (iii) an s+s' type superfluid [26] (invariant under I), and (iv) an s+s' chiral superfluid (*T* and *I* broken). The four vector types may be classified in a similar way : (i) axial vector symmetric under *T* (e.g. *a*=<s $\times$ *m*> where *s* is the spin density and *m* the magnetization), (ii) axial vector anti-symmetric under *T* (e.g. magnetization *m*), (iii) polar vector symmetric under *T* (e.g. polarization *p*) and (iv) polar vector anti-symmetric under *T* (e.g. toroidal moment [20] *t* or electric current density).

**Table 1**. Directions of the four types of vectors (*t*,*m*,*p*,*a*) induced by illumination in the four types of isotropic liquids. $I_3$ is the unit vector parallel to the wave vector *k*.

|  | *t* | *m* | *p* | *a* |
|---|---|---|---|---|
| (i) normal achiral | $I_3$ | 0 | 0 | 0 |
| (ii) normal chiral | $I_3$ | $I_3$ | 0 | 0 |
| (iii) s+s' achiral | $I_3$ | 0 | $I_3$ | 0 |
| (iv) s+s' chiral | $I_3$ | $I_3$ | $I_3$ | $I_3$ |

### III. ALTERNATVIVE THEORETICAL APPROACH TO THE PHOTOVOLTAIV EFFECT

For analysing theoretically the photovoltaic effect in a material one can assume that the electric current response of the material to a light beam depends on the electric field of the beam and on the beam propagation direction determined by the wave vector *k*. Considering, for example, a linearly polarized wave associated with the electric field amplitude *e*, wave vector *k* and phase $\phi$, the components of the induced current density *J* in the material read:

$J^i = J^i(\alpha,\beta,\gamma,\phi,I_{ph})$   (2)

where $\alpha,\beta$ are the spherical angles of *k*, $\gamma$ is the angle of *e* in the plane perpendicular to *k* (Fig. 1), and $I_{ph}$ is the beam intensity. For a static response of the material to the light beam the $\phi$-dependence disappears since the $J^i$ functions are averaged over times much larger than the period of the light wave. The $J^i$ response functions are constrained by the point-group symmetry of the crystal and by the mean symmetry of the light wave. Both constraints are taken into account using an approach in which the $J^i$ are expanded in Wigner spherical functions:

$$J^i(\mu) = \sum_{L=0}^{\infty} \sum_{n=-L}^{L} \sum_{p=-L}^{L} {}^i K_L^{pn} D_{np}^L(\mu)  \quad (3$$

Where ($\mu$) stands for the Euler angles ($\alpha,\beta,\gamma$), $D_{pn}^L(\mu)$ are spherical functions, and the coefficients ${}^iK_L^{pn}$ represent tensors of various ranks, the non-zero components of which are determined by the point-group symmetry of the crystal. When several components are permitted by symmetry the angular variation of the current density expresses complex photovoltaic effects. Furthermore, a phase transition gives rise to additional components ${}^iK_L^{pn}$ typifying the corresponding lowering of symmetry. As shown in section V, Wigner expansion ensures a direct accounting of the crystal and light wave point-group symmetries, since the functions $D_{pn}^L(\mu)$ transform as irreducible representations of both point-groups.

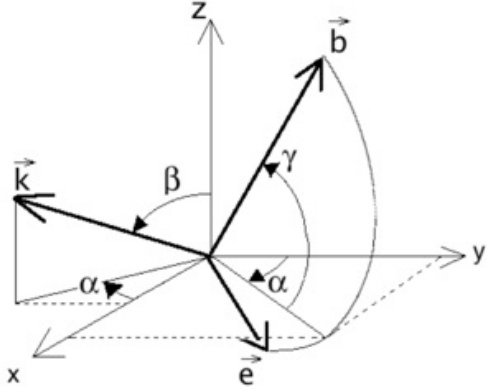

**Figure 1:** Definition of the Euler angles $\alpha\beta\gamma$ characterizing the orientation of the wave frame **e,b,k** with respect to the laboratory frame xyz.

Thus, both groups provide an independent restriction of the Wigner expansion form. We denote by *external selection* the corresponding restriction procedure involving the wave symmetry operations and denote by *internal selection* that involving crystal symmetries (details of the corresponding internal and external algebras can be found in B.Mettout [16], within another physical context). These two procedures, presented in section VI, allow us to write the most general form of the induced current response functions associated with each of the 122 magnetic point groups [17], particularly those of multiferroic materials in which both time and space reversal symmetries are usually lost [18,19].

Excitation of electrons in the conduction band of semiconductor materials can also modify their equilibrium structural or magnetic properties. Our approach applies to response functions associated, in steady states, with any physical quantity described by tensors of any rank $N$ and any symmetry with respect to time, $T$, and space, $I$, reversals. Thus, it permits accounting for a large variety of additional photo-induced phenomena. When $N=1$, we have worked out the response functions associated with (i) a polar vector antisymmetric with respect to T, which describes the *photovoltaic effect* when the vector is the electric current as well as the *phototoroidal effect* when the vector is the toroidal moment [20]; the response functions associated with (ii) a polar vector symmetric with respect to T, which describes the *optical rectification effect* [21] when the vector is the macroscopic polarization, and the response function associated with (iii) an axial vector antisymmetric with respect to T, which describes the *photomagnetic effect* [22] when the vector is the macroscopic magnetization. When $N=2$, the response functions of the dielectric and magnetic permittivity and of the elastic strain tensor (symmetric with respect to both T and I) describe the *photorefractive effect* [23] involved in non-linear optics and the *inverse photoelastic effect* [24].

Finally, the linear magnetoelectric tensor (anti-symmetric under $T$ and $I$) decribes the *photo-magneto-electric effect* [25].

## IV. MEAN SYMMETRY OF THE LIGHT BEAM

Light acts locally as an external generalized field on the crystalline and magnetic structures of illuminated materials. Indeed, the response of the crystal at macroscopic time scale is static since it involves a time average over many periods of the electromagnetic wave. Moreover, when the crystal is crossed by the wave it appears as a continuum since the wavelength is much larger than the lattice spacing. The irradiation effects on the material properties can then be evaluated by considering static and homogeneous tensor quantities in exactly the same way as when the material is submitted to static fields [25]. The onset rules of such tensors follow the same symmetry constraints in both cases and can be approached phenomenologically.

In ordered media the response functions depend on the crystal (or order parameter) point symmetry, together with the symmetry of the mean wave. Let us first analyze the linearly polarized wave symmetry group. **e** and **b** represent the wave amplitudes, **k** its wave vector and $\phi$ its phase (**E** should not be confused with **E** in equation (1)):

$E = e\, exp\, i(k.r-\omega t+\phi)$
$B = b\, exp\, i(k.r-\omega t+\phi)$

**e**, **b** and **k** are parallel to an orthonormal frame $\Sigma_\mu = I_1, I_2, I_3$ ($\mu$ represents a set of Euler angles that will be defined precisely hereafter). In contrast to (**e,b,k**), ($I_1, I_2, I_3$) are polar vectors invariant under time reversal. The mean symmetry group $G_{em}$ of the wave contains at least the group of the triplet **e,b,k**. The latter is the point group 2'mm' where 2' means the combination of time reversal $T$ with a twofold rotation axis 2 around $I_1$, and m' means the combination of $T$ with a mirror plane m normal to $I_2$. In addition, after a twofold rotation around **k**, **e** and **b** are transformed into $-e$ and $-b$. This can be compensated by a $\pi$ phase shift of $\phi$ in equation (1). Since the mean properties of the wave are independent of the phase, this rotation belongs also to $G_{em}$. One finds finally ($I$ stands for space inversion):

$G_{em}=mmm'=\{e, C_\pi(I_1)T, C_\pi(I_2)I, C_\pi(I_3)IT\} \otimes \{e, IT\}$, (3)

which is isomorphic to $D_{2h}$. $C_\pi(I_3)$ stands for the twofold rotation around the axis $I_3$. $G_{em}$ has three generators: $IT$, $C_\pi(I_3)$ and $C_\pi(I_1)T$ or, equivalently, $C_\pi(I_1)T$, $C_\pi(I_2)I$ and $C_\pi(I_3)$.

When the beam travels within a material it breaks its symmetry and, accordingly, it induces the onset of a number of physical quantities. Although $G_{em}$ does not permit the onset of any non-trivial scalar quantity (i.e. anti-symmetric with respect to time and/or space reversals, $I$ and $T$), it permits a toroidal vector (anti-symmetric under $I$ and $T$) parallel to **k** (e.g. the Poynting vector). The current density has the same symmetry properties; nevertheless no current is present when the beam travels in the vacuum because no electric charge is then available, whereas, in a conductor, an electric current can flow in the direction of the Poynting vector (photoelectric effect). Analogously, $G_{em}$ permits a second-rank tensor symmetric under $I$ and $T$ (e.g. deformation tensor or dielectric susceptibility) with non-zero diagonal components (in the frame $I_1, I_2, I_3$) and a second-rank tensor anti-symmetric under both $I$ and $T$ (e.g. magneto-electric susceptibility) with two non-zero components, $T^{13}$ and $T^{31}$. All these effects are present in any material, even in isotropic liquids. For crystalline materials many other effects can take place under light irradiation, depending on the symmetry of the material.

One can also illustrate the wave symmetry-breaking effects by considering the onset of various types of physical vectors when various types of isotropic liquids are illuminated. No calculation is then necessary to predict the directions of these vectors, as summarized in Table I: They are either forbidden by symmetry or directed parallel to $k$. The four types of isotropic liquid are (i) a normal achiral liquid (invariant under $T$ and $I$), (ii) a non-racemic mixture of chiral molecules (invariant under T), (iii) an s+s' type superfluid [26] (invariant under I), and (iv) an s+s' chiral superfluid ($T$ and $I$ broken). The four vector types may be classified in a similar way : (i) axial vector symmetric under $T$ (e.g. $a = \langle s \times m \rangle$ where $s$ is the spin density and $m$ the magnetization), (ii) axial vector anti-symmetric under $T$ (e.g. magnetization $m$), (iii) polar vector symmetric under $T$ (e.g. polarization $p$) and (iv) polar vector anti-symmetric under $T$ (e.g. toroidal moment [20] $t$ or electric current density).

**Table 1**. Directions of the four types of vectors ($t, m, p, a$) induced by illumination in the four types of isotropic liquids. $I_3$ is the unit vector parallel to the wave vector $k$.

|                     | $t$   | $m$   | $p$   | $a$   |
|---------------------|-------|-------|-------|-------|
| (i) normal achiral  | $I_3$ | 0     | 0     | 0     |
| (ii) normal chiral  | $I_3$ | $I_3$ | 0     | 0     |
| (iii) s+s' achiral  | $I_3$ | 0     | $I_3$ | 0     |
| (iv) s+s' chiral    | $I_3$ | $I_3$ | $I_3$ | $I_3$ |

## V. RESPONSE FUNCTION UNDER LINEARLY POLARIZED LIGHT
### A. Achiral and non-magnetic materials

In non-isotropic materials more intricate situations may occur, depending on both wave and crystal symmetries. One considers the interaction between crystals and linearly polarized waves independently oriented with respect to the laboratory frame $\Sigma_0 = i_1, i_2, i_3$. In order to describe the orientation of the crystal axes in the laboratory, let us attach an orthonormal frame $\Sigma_v = e_1, e_2, e_3$ to the crystal structure. One obtains both crystal ($\Sigma_\mu$) and wave ($\Sigma_v$) frames by applying to $\Sigma_0$ the two rotations defined in the laboratory frame by the sets of Euler angles $\mu = \{\alpha \beta \gamma\}$ and $v = \{\alpha' \beta' \gamma'\}$, respectively. More generally, for given sets of Euler angles, $\sigma$ and $\tau$, let us define the rotation operator $R(\tau, \sigma)$ with Euler angles $\sigma$ with respect to the frame $\Sigma_\tau$. Denoting $R(0, \sigma)$ by $R(\sigma)$ one obtains:

$$R(\tau, \sigma) = R(\tau) R(\sigma) R(-\tau), \quad (4)$$

where $R(-\tau)$ stands for $R^{-1}(\tau)$.

We are interested by the physical quantities induced by the wave illumination. These quantities are classified into tensor classes characterized by their rank $N$ (e.g. vectors when $N=1$). The induced laboratory Cartesian components $H^A = H^{a_1 a_2 ... a_N}$ of a $N^{th}$-rank tensor $H$ in the crystal depend on the orientations of both wave ($\mu$) and crystal ($v$), and on the wave amplitude and frequency ($A$ stands for the multi index $a_1, a_2, ..., a_N$):

$$H^A = \underline{H}^A + \Sigma^A(v, \mu)$$
$$H = \underline{H} + \Sigma(v, \mu), \quad (5)$$

where $\underline{H}$ is the tensor in absence of radiation field. In principle one can expand the response function $\Sigma^A$ in Wigner functions with respect to $\mu$ and $v$ separately. However, the transformation properties of $\Sigma^A$ permit us to simplify this expansion. Indeed, since $\Sigma^A$ is a tensor it obeys the covariance condition (we use the Einstein summation convention for repeated Cartesian indices $A$ and $a_i$):

$$\Sigma^A(R(\tau)v, R(\tau)\mu) = R(\tau)^{a_1}{}_{b_1} R(\tau)^{a_2}{}_{b_2} ... R(\tau)^{a_N}{}_{b_N} \Sigma^B(v, \mu), \quad (6)$$

where the set of Euler angles $\upsilon = R(\tau)\mu$ is defined by the relation $R(\upsilon) = R(\tau) R(\mu)$. Equation (6) yields:

$$\Sigma^A(v, \mu) = R(v)^A{}_B \Sigma^B(0, R(-v)\mu), \quad (7)$$

where $R(v)^A{}_B = R(v)^{a_1}{}_{b_1} R(v)^{a_2}{}_{b_2} ... R(v)^{a_N}{}_{b_N}$.

Let us expand $\sigma^A(\mu) = \Sigma^A(0, \mu)$ in Wigner spherical functions[15] $D^L_{nm}(\mu) = D^S(\mu)$, where $S$ stands for the three-index $^L_{nm}$ with $-L \leq m, n \leq +L$:

$$\sigma^A(\mu) = \sum_S {}^A K_S D^S(\mu) \quad (8)$$

The expansion coefficients ${}^A K_S$ are given by the integral

$$^A K_S = \frac{2L+1}{8\pi^2} \int \Sigma^A(0, \mu) D^S(\mu) d^3\mu \quad (9)$$

where $d^3\mu = \sin\alpha \, d\alpha \, d\beta \, d\gamma$. Since $\sigma^A$ is a real function ${}^A K_L{}^{nm}$ obeys the *reality condition*:

$${}^A K_L{}^{nm*} = (-1)^{n-m} {}^A K_L{}^{-n,-m} \quad (10)$$

The response function $\sigma^A$, which coincides with $\Sigma^A$ when the crystal axes are parallel to the laboratory axes, contains all the internal information about the crystal. It depends on its thermodynamic properties, and in particular on its symmetry group.

Equations (7,8) provide the general covariant form of the response function:

$$\Sigma^A(v, \mu) = R(v)^A{}_B \sum_{Lnm} {}^B K_L{}^{nm} D^L_{nm}(R(-v)\mu) \quad (11)$$

that can be rewritten by using basic properties of Wigner functions as:

$$\Sigma^A(v, \mu) = R(v)^A{}_B \sum_{Lnmp} {}^B K_L{}^{nm} D^L_{np}(-v) D^L_{pm}(\mu) \quad (12)$$

This expansion remains valid even for non-symmetric crystals and waves. However, restrictions in equation (12) must appear when one considers these symmetries. We will denote hereafter by *internal* and *external*[16] the crystal and wave symmetries, respectively.

The action of crystal and wave rotations on $\Sigma^A$ is immediately deduced from equation (7). Let us denote by $R^{ext}(\beta)$ the external rotation to an angle $\beta$ defined in the crystal frame, and by $R^{int}(\beta)$ the corresponding internal rotation:

$$[R^{ext}(\beta).\Sigma]^A(\nu,\mu) = \Sigma^A(\nu, R(\mu,\beta)\mu) = \Sigma^A(\nu, R(\mu)\beta)$$
$$[R^{int}(\beta).\Sigma]^A(\nu,\mu) = \Sigma^A(R(\nu,\beta)\nu, \mu) = \Sigma^A(R(\nu)\beta,\mu), \quad (13)$$

which lead to:

$$[R^{ext}(\beta).\sigma]^A(\mu) = \sigma^A(R(\mu)\beta)$$
$$[R^{int}(\beta).\sigma]^A(\mu) = R(\beta)^A{}_B \, \sigma^B(R(-\beta)\mu). \quad (14)$$

Combining equation (14) with equation (8) provides the transformation laws of the coefficients ${}^A K_S$:

$$^A K''^{np}_L = \sum_m {}^A K^{nm}_L D^L_{pm}(-\beta)$$

$$^A K'^{np}_L = R(\beta)^A_B \sum_n {}^B K^{mp}_L D^L_{mn}(-\beta) \quad (15)$$

where ${}^A K'_S$ and ${}^A K''_S$ are the Wigner coefficients of $R^{ext}(\beta).\sigma$ and $R^{int}(\beta).\sigma$, respectively. Considering, for fixed values of $A$ and $L$, that the coefficients ${}^A K''^{np}_L$ form à $(2L+1) \times (2L+1)$ matrix with indices $n$ and $p$, one can see in equations (14) that external rotations act only on the second index $p$, whereas internal rotations act on the first index $n$. More specifically, each raw of the matrix transforms under external rotations as an irreducible $L^{th}$-rank tensor (index $p$), whereas each (set of) line(s) transforms as the direct product of an irreducible $L^{th}$-rank tensor (index $n$) with a $N^{th}$-rank tensor (index $A$).

### B. Chiral and/or magnetic materials

When considering a crystal in a chiral phase the description must be refined. Indeed, such a crystal can appear under two distinct domain types, right-handed and left-handed, transforming one into another by applying the space reversal symmetry $I$, or a combination of $I$ with a rotation. For a right-handed crystal we choose, as in the previous section, a direct (right-handed) frame $e_1,e_2,e_3$. For a left-handed crystal we choose a left-handed frame $f_1,f_2,f_3$. When the left-handed crystal is obtained from the right-handed crystal by applying $I$ alone, one conventionally defines $f_m = -e_m$.

For characterizing the orientation of the crystal one has now to introduce a chiral index $\kappa_I$, with $\kappa_I=1$ for right crystals and $\kappa_I=-1$ for left crystals. The complete crystal orientation is then described by the couple $(\kappa_I,\nu)$. $\nu$ represents the Euler angles transforming $\Sigma_0 = i_1,i_2,i_3$ into $e_1,e_2,e_3$ for right-handed frames, as in the previous section, and the Euler angles transforming $I.\Sigma_0 = -i_1,-i_2,-i_3$ into $f_1,f_2,f_3$ for left-handed frames.

Similarly, for magnetic materials one has to consider two types of magnetic domains transforming one into another by the time reversal symmetry $T$. The magnetic index $\kappa_T=1$ for one type of domain, and $\kappa_T=-1$ for the other type. Since the frames are all symmetric under time reversal, there is no need to define specific basis for $\kappa_T=-1$

Thus, in the general case the response function depends additionally on the two previous indices:

$$\Sigma^A = \Sigma^A(\kappa_I, \kappa_T, \nu, \mu), \quad (16)$$

and the Wigner coefficients in equation (8) depend on $\kappa_I$ and $\kappa_T$: ${}^A K_S(\kappa_I, \kappa_T)$.

In order to present a unified formalism we will use also equation (16) for the response function of non-magnetic and achiral materials. In the case of an achiral material we permit now to attach right-handed as well as left-handed frames to the crystal. They transform ones into others by space inversion and mirror planes of the material. For instance, when the symmetry group contains $I$ we can attach $e_1,e_2,e_3$ as well as $-e_1,-e_2,-e_3$. Thus, since using any type of frame yields the same response functions, $\Sigma^A$ must be independent of $\kappa_I$. Analogously in a material containing $T$ in its magnetic point group the response function cannot depend on $\kappa_T$.

Pure rotations have no effect on $\kappa_I$ and $\kappa_T$, whereas $I$ transforms $\kappa_I$ without changing $\kappa_T$, and reciprocally for $T$:

$$I \, \kappa_I, \kappa_T, \nu = -\kappa_I, \kappa_T, \nu$$
$$T \, \kappa_I, \kappa_T, \nu = \kappa_I, -\kappa_T, \nu$$
$$IT \, \kappa_I, \kappa_T, \nu = -\kappa_I, -\kappa_T, \nu. \quad (17)$$

In addition, the tensor $\Sigma$ can be either symmetric or anti-symmetric with respect to both space and time reversals. Let us define the corresponding chiral and magnetic tensor indices $\tau_I$ and $\tau_T$: $\tau_I=1$ for tensors symmetric under $I$, and $\tau_I=-1$ for anti-symmetric tensors. Similarly, $\tau_T=\pm 1$ for tensors symmetric or anti-symmetric under time reversal. The type of a given tensor is thus defined by its rank $N$ together with its two indices: $[N,\tau_I,\tau_T]$. For instance, the induced current density of the photovoltaic effect is a tensor of type $[1,-1,-1]$, i.e. a polar vector anti-symmetric under time reversal.

The response function $\Sigma^A$ must also be covariant under $I$ and $T$. When applying $I$ or $T$ to both crystal and wave the tensor $\Sigma^A$ must remain unchanged up to a factor $\tau_I$ or $\tau_T$. Let us denote by $I^{ext}$ and $T^{ext}$ the actions of $I$ and $T$ on $\Sigma_\nu$ (notice that applying $I$ and $T$ on the triplet $e,b,k$ transforms a right-handed frame $\Sigma_\nu$ into a right-handed frame). The covariance conditions for $I$ and $T$, analogous to equation (13) for rotations, read:

$$[I^{ext}\sigma]^A(\kappa_I,\kappa_T,\mu) = \tau_I \, \sigma^A(-\kappa_I,\kappa_T,\mu)$$
$$[T^{ext}\sigma]^A(\kappa_I,\kappa_T,\mu) = \tau_T \, \sigma^A(\kappa_I,-\kappa_T,\mu). \quad (18)$$

Let us emphasize the difference between equations (18) and a way to approach covariance stating that reversing a chiral domain yields multiplying $\sigma^A$ by $\tau_I$ (e.g. in two opposite chiral domains polarization is reversed and magnetization conserved). This would be true if the system was isolated. When the system is submitted to an external symmetry-breaking field (or wave) this naive statement must be replaced with equations (18).

### VI. SELECTION PROCEDURES
### A. External selection

In order to evaluate the action of $I^{ext}$ and $T^{ext}$ on $\mu$ in equations (18), let us recall that the wave group is mmm' (see equation (3)). Thus, applying $I$ on the wave has the same effect as applying the twofold rotation $C_\pi(I_2)$ around

$I_2$ in the wave frame. Similarly, applying $T$ is equivalent to applying $C_\pi(I_1)$:

$$I^{ext} \mu = C_\pi(I_2) \mu$$
$$T^{ext} \mu = C_\pi(I_3) \mu. \quad (19)$$

One can use equations (19) to express the constraints imposed on the response function form by the symmetry of the wave. Indeed, applying anyone of the generators $C_\pi(I_2)I$, $C_\pi(I_3)$ and $C_\pi(I_1)T$ of $G_{em}$ cannot modify the response of the crystal, yielding three constraints restricting the expansion of $\Sigma^A$:

Equations (19) show that applying $I$ on the wave has the same effect as applying the rotation $C_\pi(I_1)$ characterized by Euler angles $\{0\pi\pi\}$, that is $I$ transforms $\Sigma^A$ into $\Sigma^A(\kappa_I,\kappa_T,\nu,R(\mu)\{0\pi\pi\})$. Combining this result with equation (18) yields the first internal constraint:

$$\sigma^A(\kappa_I,\kappa_T,R(\mu)\{0\pi\pi\}) = \tau_T \sigma^A(\kappa_I,-\kappa_T,\mu) \quad (20)$$

A similar reasoning yields the constraint associated with the two other generators generators:

$$\sigma^A(\kappa_I,\kappa_T,R(\mu)\{0\pi 0\}) = \tau_I \sigma^A(-\kappa_I,\kappa_T,\mu).$$
$$\sigma^A(\kappa_I,\kappa_T,R(\mu)\{00\pi\}) = \sigma^A(\kappa_I,\kappa_T,\mu) \quad (21)$$

Constraints (20) and (21) cancel a number of Wigner coefficients in equation (8). We denote by *external selection* the corresponding procedure. It yields from equations (20) the first external selection rule:

$$^A K_L^{nm}(\kappa_I,-\kappa_T) = (-1)^L \tau_T {}^A K_L^{n,-m}(\kappa_I,\kappa_T) \quad (22)$$

The last external selection rules, arising from equation (21), read:

$$^A K_L^{n-m}(-\kappa_I,\kappa_T) = (-1)^{L+m} \tau_I {}^A K_L^{nm}(\kappa_I,\kappa_T),$$
$$^A K_L^{n,2m+1}(\kappa_I,\kappa_T) = 0 \quad (23)$$

The external selection does not involve the index $A$, so that it applies in the same way on each tensor component. In particular, it cannot prevent the onset of any specific component of a given tensor, but only restrict its angular dependence.

The T-covariance condition in Eq.(18) and, accordingly, the T-selection rule shown in Eq.(22), assume the reversibility of the considered process. Both light absorption and electric conductivity are irreversible phenomena, so that Eq.(22) can only be considered as a first approximation, valid at very low beam intensity. Thus, one has to distinguish in the Wigner expansion given by Eq.(8), the terms which obey Eq.(22) from those which do not. The formers are called non-dissipative terms, whereas the latters are dissipative terms [1]. They all vanish when the beam intensity tends to zero, however the dissipative terms should vanish more strongly than the non dissipative ones.

### B. Internal selection

The magnetic point group of the crystal restricts also the form of the response function: $\sigma^A$ is not modified when one applies internal transformations belonging to the magnetic point group $G_{cr}$. The action of these symmetries on the response function is defined by equations (14,15,17). The corresponding constraints define the internal selection procedure and give rise, for each generator of $G_{cr}$, to an internal selection rule analogous to the external selection rules presented in equations (22,23).

For instance, if $G_{cr}$ contains space and time reversals then the constraints read:

$$\sigma^A(\kappa_I,\kappa_T,\mu) = \sigma^A(-\kappa_I,\kappa_T,\mu)$$
$$\sigma^A(\kappa_I,\kappa_T,\mu) = \sigma^A(\kappa_I,-\kappa_T,\mu), \quad (24)$$

yielding internal selection rules:

$$^A K_L^{nm}(-\kappa_I,\kappa_T) = {}^A K_L^{nm}(\kappa_I,\kappa_T)$$
$$^A K_L^{nm}(\kappa_I,-\kappa_T) = {}^A K_L^{nm}(\kappa_I,\kappa_T). \quad (25)$$

If $G_{cr}$ contains a pure rotation, $R^{int}(\beta)$, the constraint and the selection read (see equations (14,15)):

$$R(\beta)^A{}_B \sigma^B((\kappa_I,\kappa_T,R(-\beta)\mu) = \sigma^A(\kappa_I,\kappa_T,\mu)$$
$$R(\beta)^A{}_B \sum_m {}^B K_L^{mn}(\kappa_I,\kappa_T) D^L_{mp}(-\beta) = {}^A K_L^{pn}(\kappa_I,\kappa_T). \quad (26)$$

The constraint and the selection rule associated with a general internal symmetry combining a rotation with $I$ and/or $T$, can be found on following the same method.

In practical calculations equations (26) are not always easy to handle due to the presence of the multi-indices $A$ and $B$. For simplicity, it is preferable to work with irreducible tensors containing $2L+1$ components instead of $N^2$ components for the associated reducible tensor. For instance, a second rank irreducible tensor is defined as the symmetric traceless part (with five independent components) of the associated reducible tensor (with nine components).

Let us consider now that $\sigma$ is irreducible. It can be decomposed onto a *standard basis* of tensors $B_M$ (transforming under rotations like the spherical harmonics $Y_N^M$) with $-N \leq M \leq +N$:

$$\sigma(\kappa_I,\kappa_T,\mu) = \sum_{M=-N}^{N} \tilde{\sigma}^M(\kappa_I,\kappa_T,\mu) B_M \quad (27)$$

The spherical tensor components $\tilde{\sigma}^M$ are complex functions transforming under rotation as:

$$\left(R^{int}(\beta).\tilde{\sigma}\right)^M(\kappa_I,\kappa_T,\mu) = \sum_{P=-N}^{N} \tilde{\sigma}^P(\kappa_I,\kappa_T,R(-\beta)\mu) D^N_{PM}(\beta) = \sum_{P=-N}^{N} \sum_{L=0}^{\infty} \sum_{mnq=-L}^{L} {}^P \tilde{K}_L^{mn}(\kappa_I,\kappa_T) D^L_{mq}(-\beta) D^L_{qn}(\mu) D^N_{PM}(\beta)$$

$$(28)$$

Thus, the selection rules presented in equations (26) can be rewritten as:

$$\sum_{P=-N}^{N} \sum_{L=0}^{\infty} \sum_{m=-L}^{L} {}^P \tilde{K}_L^{mn} D^L_{mq}(\beta) D^N_{PM}(-\beta) = {}^M \tilde{K}_L^{qn} \quad (29)$$

that are simpler than equations (26) because M and P are integer numbers. equations (24,25) are not modified when one replaces Cartesian components $\sigma^A$ with spherical components $\tilde{\sigma}^M$. Thus, for each internal generator equations (24,25,29) provide the internal selection rules giving the most general form of the response function compatible with the crystal symmetries.

An isotropic system does not permit any tensor of rank larger than zero. equation (29) shows that, for fixed $N$, $L$ and $n$, the $(2N+1) \times (2L+1)$ coefficients ${}^P \tilde{K}_L^{mn}$ transform as the direct product $\Gamma^{(N)} \otimes \Gamma^{(L)}$ of irreducible representations of the isotropic rotation group $SO(3)$. This product contains one scalar only when $N=L$. Since in an isotropic system only scalar quantities can spontaneously be present at equilibrium, one sees that for a given $N$ only one value of $L$ survives in expansion (8).

Along the same way, in a system with uniaxial symmetry group SO(2) (say, around the Oz axis) equation (29) shows that only the coefficients ${}^P\widetilde{K}_L^{m,n}$ with $m=P$ can survive.

## VII. PHOTO-MAGNETO-ELECTRIC EFFECTS IN LiNbO$_3$
### A. Second order Photovoltaic effect

In the photovoltaic effect there is no spontaneous current ($H=0$ in equation (5)) and the response function $\boldsymbol{\sigma}$, namely the light-induced electric current, is a tensor of type [1,-1,-1]. Thus, the multi-index $A$ in equation (5) becomes a single integer index: $a=1,2,3$. Writing explicitly $\boldsymbol{\sigma}$ and the Wigner coefficients versus $\kappa_I$ and $\kappa_T$:

$$\boldsymbol{\sigma} = \boldsymbol{\sigma}(0) + \kappa_T\,\boldsymbol{\sigma}(1) + \kappa_I\,\boldsymbol{\sigma}(2) + \kappa_I\,\kappa_T\,\boldsymbol{\sigma}(3)$$
$${}^aK_L^{n,m}(\kappa_I, \kappa_T) = {}^aK(0)_L^{n,m} + \kappa_T\,{}^aK(1)_L^{n,m}$$
$$+ \kappa_I\,{}^aK(2)_L^{n,m} + \kappa_I\,\kappa_T\,{}^aK(3)_L^{n,m}, \quad (30)$$

the reality condition and the external selection presented in equations (10,23) become:
$${}^aK(s)_L^{n,m} = (-1)^{n-m}\,{}^aK(s)_L^{-n,-m}*$$
$${}^aK(s)_L^{n,2m+1} = 0$$
$${}^aK(s)_L^{n,m} = \phi_s\,(-1)^{L+l+m}\,{}^aK(s)_L^{n,-m}, \quad (31)$$

where $s=0,1,2,3$ and $\phi_s=1$ for $s=0,1$ and $\phi_s=-1$ for $s=2,3$. If one restricts the expansion of $\boldsymbol{\sigma}$ in equation (8) to $L\leq 2$, and writes:

$$\boldsymbol{\sigma}(s) = \boldsymbol{\sigma}_{(0)}(s) + \boldsymbol{\sigma}_{(1)}(s) + \boldsymbol{\sigma}_{(2)}(s), \quad (32)$$

where $\boldsymbol{\sigma}_{(L)}(s)$ contains the $(2L+1)^2$ Wigner functions $D^L_{mp}(\mu)$ occurring in equation (8), then one finds finally (with $\mu=\{\alpha,\beta,\gamma\}$):

$\boldsymbol{\sigma}_{(0)}(s) = A_s\,(1-\mu_s)$
$\boldsymbol{\sigma}_{(1)}(s) = \{B_s\cos\alpha + C_s\sin\alpha\}\sin\beta\,\mu_s + D_s\cos\beta\,\mu_s$
$\boldsymbol{\sigma}_{(2)}(s) = E_s\,(1-3\cos^2\beta)(1-\mu_s)$
$\quad + \{F_s\cos\alpha + G_s\sin\alpha\}\sin 2\beta\,(1-\mu_s)$
$\quad + \{H_s\cos 2\alpha + I_s\sin 2\alpha\}\sin^2\beta\,(1-\mu_s)$
$\quad + J_s\sin^2\beta\sin 2\gamma\,\mu_s + K_s\sin^2\beta\cos 2\gamma(1-\mu_s)$
$\quad + L_s\sin\beta\{\cos(\alpha+2\gamma)\cos^2\beta/2+\cos(\alpha-2\gamma)\sin^2\beta/2\}\,\mu_s$
$\quad + M_s\sin\beta\{\sin(\alpha+2\gamma)\cos^2\beta/2+\sin(\alpha-2\gamma)\sin^2\beta/2\}\,\mu_s$
$\quad + N_s\sin\beta\{\cos(\alpha+2\gamma)\cos^2\beta/2-\cos(\alpha-2\gamma)\sin^2\beta/2\}(1-\mu_s)$
$\quad ,+ O_s\sin\beta\{\sin(\alpha+2\gamma)\cos^2\beta/2-\sin(\alpha-2\gamma)\sin^2\beta/2\}(1-\mu_s)$
$\quad + P_s\,\{\cos 2(\alpha+\gamma)\cos^4\beta/2-\cos 2(\alpha-\gamma)\sin^4\beta/2\}\,\mu_s$
$\quad + Q_s\,\{\sin 2(\alpha+\gamma)\cos^4\beta/2-\sin 2(\alpha-\gamma)\sin^4\beta/2\}\,\mu_s$
$\quad + R_s\,\{\cos 2(\alpha+\gamma)\cos^4\beta/2+\cos 2(\alpha-\gamma)\sin^4\beta/2\}(1-\mu_s)$
$\quad + S_s\{\sin 2(\alpha+\gamma)\cos^4\beta/2+\sin 2(\alpha-\gamma)\sin^4\beta/2\}(1-\mu_s),\quad (33)$

where $\mu_s=1$ when $s=0,1$ and $\mu_s=0$ when $s=2,3$ if one takes into account non dissipative as well as dissipative terms in the Wigner expansion. If one takes only into account non-dissipative terms, the coefficients with $s=1$ or 2 must be neglected. Since $A_s$, $B_s$ ... $S_s$ are vector coefficients (proportional to the non-vanishing ${}^aK(s)_L^{n,m}$), equations (32) show that up to the second order the response function depends on 114 real parameters. If the magnetic point group of the crystal is reduced to the identity, then all these coefficients are independent. However, for non-trivial magnetic point groups the internal selection diminishes the number of independent coefficients.

Let us work out the response function in LiNbO$_3$ as a first illustration of our approach. The photovoltaic effect in this material was first observed in 1969 and shown to take its origin in the excitation of impurity electrons into the conduction band, and in their induced motion by the internal electric field generated by the spontaneous crystal polarization. Indeed, LiNbO$_3$ is a ferroelectric polar and non-magnetic material with point symmetry 3m1'. We set the Oz direction of the sample frame along the three-fold symmetry axis, and Ox normal to one of the three mirror planes.

Since $G_{cr}$ contains time reversal $T$, equation (24) shows $\boldsymbol{\sigma}$ does not depend on $\kappa_T$, so that $\boldsymbol{\sigma}(1)=\boldsymbol{\sigma}(3)=0$. Despite the achiral character of 3m1', $\boldsymbol{\sigma}$ depends on $\kappa_I$ because the inversion $I$ does not belong to the group. Changing $\kappa_I$ amounts to reverse the spontaneous polarization of the crystal.

On the other hand, equations (25,26) for the three-fold internal rotation and the mirror plane cancel a number of vector coefficients in equations (33). Finally, the second-order photovoltaic complete response functions (with dissipative terms) in LiNbO$_3$ read (in the crystal frame xyz):

$$\boldsymbol{\sigma}(0) = \begin{pmatrix} -B_{0y}\sin\beta\sin\alpha \\ B_{0y}\sin\beta\cos\alpha \\ D_{0z}\cos\beta \end{pmatrix}$$

$$+O_{0x}\begin{pmatrix} \cos^4\frac{\beta}{2}\sin 2(\gamma+\alpha)-\sin^4\frac{\beta}{2}\sin 2(\alpha-\gamma) \\ \cos^4\frac{\beta}{2}\cos 2(\gamma+\alpha)-\sin^4\frac{\beta}{2}\cos 2(\alpha-\gamma) \\ 0 \end{pmatrix}$$

$$\boldsymbol{\sigma}(2) = \begin{pmatrix} -F_{2y}\sin 2\beta\sin\alpha + H_{2x}\sin^2\beta\cos 2\alpha \\ F_{2y}\sin 2\beta\cos\alpha - H_{2x}\sin^2\beta\sin 2\alpha \\ A_{2z} + E_{2z}(1-3\cos^2\beta) + K_{2z}\sin^2\beta\cos 2\gamma \end{pmatrix}$$

$$+O_{2x}\sin\beta\begin{pmatrix} \cos^2\frac{\beta}{2}\sin(\alpha+2\gamma) - \sin^2\frac{\beta}{2}\sin(\alpha-2\gamma) \\ -\cos^2\frac{\beta}{2}\cos(\alpha+2\gamma) + \sin^2\frac{\beta}{2}\cos(\alpha-2\gamma) \\ 0 \end{pmatrix}$$

$$+S_{2x}\begin{pmatrix} \cos^4\frac{\beta}{2}\sin 2(\gamma+\alpha) + \sin^4\frac{\beta}{2}\sin 2(\alpha-\gamma) \\ \cos^4\frac{\beta}{2}\cos 2(\gamma+\alpha) + \sin^4\frac{\beta}{2}\cos 2(\alpha-\gamma) \\ 0 \end{pmatrix}$$

$$\boldsymbol{\sigma}(1) = \boldsymbol{\sigma}(3) = \mathbf{0} \quad (34)$$

which depend on only eleven independent real phenomenological coefficients: $B_{0y},..,S_{2x}$.

One can notice that the terms proportional to $D_{0z}$ and $A_{2z}$ are parallel to the spontaneous polarization and to the internal electric field (along Oz). The first term ($D_{0z}$) is independent of the sense of the polarization. It corresponds to a relativistic effect where the current is provoked by the radiation pressure: Reversing the beam direction reverses the current which vanishes when the beam arrives on the crystal perpendicularly to Oz. Oppositely, in the second term ($A_{2z}$) the direction of the current is reversed when $\kappa_I$ changes from 1 to -1, that is when the spontaneous internal electric field is reversed, whereas its sense is independent of the sense of the beam and its intensity independent of the polarization of the wave. This current is provoked by the internal field induced motion of the excited electrons. Thus, the corresponding coefficients should obey the relations: $D_{0z}<0$, $A_{2z}>0$ (if one chooses the $\kappa_I=+1$ domain with spontaneous polarization $P_z>0$), and $|D_{0z}|<<A_{2z}$.

## B. First-order photo-magnetic, photo-toroidal and optic rectification effects

In multiferroic materials, light beams generate the onset of various kinds of measurable vector quantities: Polarization $P$ (type [1,-1,1]), magnetization $M$ (type [1,1,-1]) and toroidal moment $T$ (type [1,-1,-1]). The corresponding first-order covariant complete response functions compatible with the external selection are given by:

$\sigma_P(s) = A_s (1-\mu_s) + B_s \mu_s \cos\beta$
$+ C_s \mu_s \sin\beta \cos\alpha + D_s \mu_s \sin\beta \sin\alpha$
$\sigma_M(s) = A_s' \mu_s + B_s' (1-\mu_s) \cos\beta$
$+ C_s' (1-\mu_s) \sin\beta \cos\alpha + D_s' (1-\mu_s) \sin\beta \sin\alpha$
$\sigma_T(s) = A_s'' (1-\mu_s) + B_s'' \mu_s \cos\beta$
$+ C_s'' \mu_s \sin\beta \cos\alpha + D_s'' \mu_s \sin\beta \sin\alpha$ (35)

$\sigma_T(s)$ has the same form as the photovoltaic response [Eq.(33)] since both vectors have the same type [1,-1,-1] (the coefficients $A_s$... have different meanings in Eqs.(33) and (35)). In addition, on sees in Eq.(35) that $T$ and $P$ exhibit the same behavior. However, when considering only the non-dissipative terms, which are dominant at weak beam intensity, the wo response functions become distinct, because only the terms with $s=1,2$ survive in the expansion (35) for $P$, and $s=0,3$ for $T$.

In LiNbO$_3$ the additional internal conditions yield:

$$\sigma_P(0) = \begin{pmatrix} C_{0x}\sin\beta \cos\alpha \\ -C_{0x}\sin\beta \sin\alpha \\ B_{0z}\cos\beta \end{pmatrix}$$

$$\sigma_P(2) = \begin{pmatrix} 0 \\ 0 \\ A_{2z} \end{pmatrix}$$

$$\sigma_M(0) = \begin{pmatrix} 0 \\ 0 \\ A'_{0z} \end{pmatrix}$$

$$\sigma_M(2) = \begin{pmatrix} D'_{2x}\sin\beta \sin\alpha \\ D'_{2x}\sin\beta \cos\alpha \\ B'_{2z}\cos\beta \end{pmatrix}$$

$\sigma_T(s) = \sigma_P(s)$ (36)

The polarization direction ($e$) of a wave propagating along Oz ($\beta=0$) is given by the angle $\alpha+\gamma$. In this case neither polarization nor magnetization onset in the xy plane. In addition, rotating $e$ has no effect on the three vectors. When the beam is normal to the spontaneous crystal polarization ($\beta=\pi/2$) only components parallel to Oz are generated by the illumination, and they depend on the wave polarization. Rotating $e$ yields a rotation of $P$, $T$ and $M$ on circles with radii vanishing when the beam direction is aligned with the spontaneous crystal polarization.

## C. First-order photo-elastic and photo-magneto-electric effects

Light-induced structural modifications (inverse photo-elastic effect [24]) are described by the irreducible traceless part of the second-rank strain tensor (type [2,1,1]). At the first order ($L\leq1$) the external selection yields the following form for the corresponding response function in the crystal frame:

$\sigma(s) = A_s \mu_s + (1-\mu_s) B_s \cos\beta$
$+ (1-\mu_s) \sin\beta \{C_s \cos\alpha + D_s \sin\alpha\}$, (37)

where $A_s, \ldots D_s$ are constant $3\times 3$ symmetric matrices. Since LiNbO$_3$ symmetry group contains time reversal, the terms with $s$ odd are forbidden by internal selection so that, to the first order, the deformation takes the form:

$$\sigma(0) = A_0 \begin{pmatrix} -1 & 0 & 0 \\ 0 & -1 & 0 \\ 0 & 0 & 2 \end{pmatrix}$$

$$\sigma(2) = B_0 \cos\beta \begin{pmatrix} -1 & 0 & 0 \\ 0 & -1 & 0 \\ 0 & 0 & 2 \end{pmatrix}$$

$$+ B_0 \sin\beta \begin{pmatrix} \sin\alpha & \cos\alpha & 0 \\ \cos\alpha & -\sin\alpha & 0 \\ 0 & 0 & 0 \end{pmatrix}$$

$\sigma(1) = \sigma(3) = 0$ (38)

The dielectric and magnetic susceptibilities have the same type as the deformation tensor and, accordingly, satisfy the same laws (equations (37,38)) with distinct values of the coefficients $A_s\ldots$

Under low applied electric field the system acquires a magnetization proportional to the field: $M=\chi E$, where the magneto-electric susceptibility $\chi$ is a second-rank tensor of type [2,-1,-1]. It vanishes in LiNbO$_3$ because time reversal is present in its magnetic point group. Under light illumination the induced symmetry breakdown generates non-zero components of $\chi$. The external selection provides the corresponding first-order response function:

$\sigma(s) = A_s (1-\mu_s) + \mu_s B_s \cos\beta$
$+ \mu_s \sin\beta \{C_s \cos\alpha + D_s \sin\alpha\}$, (39)

where $A_s, B_s, C_s$ and $D_s$ are constant $3\times 3$ (non-necessarily symmetric) matrices. In LiNbO$_3$ it takes the form:

$$\sigma(0) = C_0 \sin\beta \begin{pmatrix} \cos\alpha & -\sin\alpha & 0 \\ -\sin\alpha & \cos\alpha & 0 \\ 0 & 0 & 0 \end{pmatrix} + B_0 \cos\beta \begin{pmatrix} 0 & 1 & 0 \\ -1 & 0 & 0 \\ 0 & 0 & 0 \end{pmatrix}$$

$$+C'_0 \sin\beta \begin{pmatrix} 0 & 0 & -\sin\alpha \\ 0 & 0 & -\cos\alpha \\ \sin\alpha & \cos\alpha & 0 \end{pmatrix}$$

$$\sigma(2) = A_2 \begin{pmatrix} 0 & 1 & 0 \\ -1 & 0 & 0 \\ 0 & 0 & 0 \end{pmatrix}$$

$$\sigma(1) = \sigma(3) = 0 \qquad (40)$$

The group 3m1' describes a non-magnetic phase and leads to rather simple effects. We have treated more refined photovoltaic effects in a system exhibiting magnetic and non-magnetic phases, $KBiFe_2O_3$, in Ref 27 on using the same methods. At the magnetic transition a number of response coefficients take non-zero values and vary with temperature in the ordered phase as the critical order parameter.

## VIII. SUMMARY AND CONCLUSION

Illumination by a linearly polarized light beam modifies the equilibrium and stationary tensors associated with any measurable quantity in crystals. We have worked out the selection rules permitting, for all tensors and all magnetic point groups, determination of the angular dependence of the corresponding response functions. The mathematical framework of Wigner functions proves to be particularly well suited for this task. Circular and elliptic wave polarizations can be treated using the same approach. They will lead, however, to distinct external selection rules. Let us note that our approach is valid only for wavelength much larger than the lattice spacing. Using X-rays would necessitate taking into account the whole magnetic space group, instead of only the point group.

Considering a vector anti-symmetric under application of time and space reversals has permitted calculating the response function expansion of the photovoltaic effect up to the second order in the Wigner functions. The three components of the induced electric current are calculated vs. the Euler angles describing the direction and polarization of the incident wave. Its symmetry-breaking effects are illustrated by the onset of various components of the current as well as by their complex angular dependence.

We have successfully applied the method to the analysis of the photovoltaic, photorefractive, photo-magnetic, photo-elastic, photo-magneto-electric, photo-toroidal and optic rectification effects. For each effect it gives general formulas valid for any material, because it results only from the external selection procedure, that is from the wave symmetry alone. The corresponding response functions depend on numerous phenomenological coefficients. Although the previous effects depend only on first- and second-rank tensors, higher-rank tensors, describing for instance elastic response or optical activity, can be calculated following the same procedure. We have no means to predict the intensity dependence of these coefficients, unless in the case of their cancellation when $I_{ph} \to 0$. Energetic arguments repeatedly used in the literature claiming that they should depend linearly on the intensity (see Eq.(1)) correspond only to a first approximation.

We have applied the previous expansions to the ferroelectric photovoltaic material $LiNbO_3$ characterized by its high point symmetry group 3m1'. Internal selection restrains the response function expansion, which diminishes drastically the number of independent response coefficients. In particular, they do not depend on the magnetic index $\kappa_T$, removing half of the coefficients, since the group of $LiNbO_3$ contains time reversal. We will publish elsewhere more refined examples of multiferroic materials for which the response functions are more intricate due to both time and space reversal breakdowns.

We have not attempted to compare our predictions with experimental results in order to keep the self-contained character of this article and because few data are presently available in the literature concerning angle-dependent photo-effects (except for the light polarization direction). Indeed, this dependence has surely no consequence for solar-cell energetic efficiency. However, all these effects should be useful for probing multiferroic material properties. In particular, they provide simple means for refining magnetic point groups. Finally, let us notice that equations (10,22,23,25,26,29) provide the form of any response function together with a more natural way for describing the time reversal breakdown as compared to the conventional power expansion.